\documentclass[twocolumn,showpacs,preprintnumbers,amsmath,amssymb]{revtex4}
\usepackage{graphicx}
\usepackage{dcolumn}
\usepackage{bm}

\begin{document}

\title{Mechanisms for the superconducting state from a one-particle derivation of the BCS gap equations.}

\author{T. Jarlborg}

\affiliation{
DPMC, University of Geneva, 24 Quai Ernest-Ansermet, CH-1211 Geneva 4,
Switzerland
}


\begin{abstract}

The BCS results for the superconducting gap $\Delta$ and $T_C$ are
obtained from a one-particle model. Superconductivity appears when 
the electronic energy gains of the band structure
surpass the energy needed for atomic vibrations or magnetic moment oscillations. 
The vibration/oscillation
amplitudes determine the superconducting gap, 
and the Fermi surface is important for the q-dependence.
This permits for complementary interpretations
of the parameters for superconductivity and modeling of
density-of-state effects. It also makes the superconducting
mechanism less exotic.

\end{abstract}


\maketitle

\section{Introduction.}


The Bardeen-Cooper-Schrieffer (BCS) theory is the basis for a microscopic 
understanding of superconductivity \cite{bcs}.
It has, together with band theory, been successful for an understanding of the correlation 
between the superconducting (SC)
$T_C$ and material parameters like phonon frequency, electron-phonon coupling, $\lambda$, and
the electronic density-of-states (DOS) at the Fermi energy, $N(E_F)$ \cite{papa,arb}.
The good understanding of the "gap ratio", the ratio between the SC 
gap, $\Delta$, at $T$=0 and 
$k_BT_C$, is another
example of the success of the BCS theory. 
The observation of isotope effects are often in good agreement with the
predictions from BCS, which confirms the dependence on lattice stiffness.
Although the variations of $T_C$ with pressure can be complex they are
usually understood from lattice hardening and the behavior of the matrix elements
for electron-phonon coupling \cite{buz}.
However, there are complications \cite{hir}; Competition with magnetism and electron-electron correlation
is reducing $T_C$. 
High-$T_C$ cuprates are not understood, since their $\lambda$'s are not very large.
The isotope effect is sometimes very complex, even in elementary metals.
Other shortcomings might be that the many-body BCS formula does not allow for 
an intuitive comprehension of the superconducting mechanism and there
is no direct explanation of the Meissner effect \cite{hir}.

Here is presented a simple one-electron like derivation of the BCS gap equation.
Many assumptions are the same as in BCS, but some interpretations are different.
The many-body formulation of BCS gap equation is based on exchange of virtual phonons and the
SC state appears "when the energy difference between the electrons
involved is less than the phonon energy" \cite{bcs}.
In the present one-particle derivation the gain in electronic energy
is caused by dynamic changes of the band structure.
A one-particle band will have a
gap because of a periodic potential perturbation, as in the appearance of
a gap for semiconductors \cite{zim}. A spontaneous lattice vibration creates
the perturbation and the electronic states at $\vec{k}$
and -$\vec{k}$ are affected equally (but there is no exchange of phonons). 
Also states on nearby k-points are involved as can be found from the band dispersion. Only
phonons which modifies electronic states around the Fermi surface (FS) are of interest,
since changes of the bands far from the Fermi energy ($E_F$) make no change in the
total kinetic energy. This puts a constraint on the $q$-vectors of the phonons.
The difference between the present derivation and original BCS is that the energy gain 
is calculated from all one-particle
states near $E_F$, and the SC state appears as soon as this gain is larger
than the vibrational energy.  
This allows for further insight to the mechanism of SC pairing,
since the parameters can be understood in a simple way from the band structure. 

\section{One-electron model at T=0.}

Phonons and spin fluctuations in the normal (non-SC) state are excited thermally following the
thermal Bose-Einstein occupation, $g(T,\omega)$, of the phonon- or spin wave DOS,
$F(\omega)$ or $F_m(\omega)$, respectively. 
The averaged atomic displacement amplitude for phonons, $u$, can be calculated 
as function of $T$ \cite{zim,grim}. Approximate results
for Debye like spectra make $u^2 \rightarrow 3\hbar\omega_D/2K$ at low $T$
and $3 k_BT/K$ at high $T$, where $K$ is an effective force constant, which can be calculated from
$K=M_A\omega^2 = d^2E/du^2$ ($E$ is the total energy), where $M_A$ is an atomic mass. The corresponding
relations for averaged fluctuation amplitudes of the magnetic moments, $m$,
are the same, but without the polarization factor 3 and with replacement of $K$ 
with $K_m = d^2E/dm^2$, which is constant
for harmonic dependence of the total energy as function of $m$ \cite{tjfe}.
Thus, for the normal state, one can estimate $u$ and $m$ at any given $T$ from 
these relations, if the
'force' factors $K$ and $K_m$ can be calculated or are known from experiment.


Phonons and spin waves have an influence on the electronic state and its DOS \cite{fesi},
and may cause a pseudo gap close to $E_F$ in the normal state of high-$T_C$ copper oxides \cite{tj1}.
Phonons make a periodic potential perturbation along a chain of atoms,
\begin{equation}
 V(x) = V_0 e^{-i \vec{x}\cdot\vec{q}}
\end{equation}
if the phonon propagation is along $\vec{x}$ with wave vector $\vec{q}$ \cite{zim,tj1}.
A spin wave makes an analogous perturbation
within the spin polarized part of the potential. The potentials for opposite spins
are the same except for a phase shift of $\pi$. The result is an anti-ferromagnetic
(AFM) spin configuration with wave length given by $2\pi/q$.
Phonons and spin waves can be considered separately, but several works
have shown that an important spin-phonon coupling (SPC) exists in the cuprates,
which also can explain many of their unusual properties \cite{tj6,tj7,tj8,egami}.

The following development is based on phonon excitations, but later 
it will be seen how things will change with spin waves and SPC.

At very low $T$ there are practically no occupation of phonons. The Fermi-Dirac
occupation $f(\epsilon,T) = 1/(e^{\epsilon/k_BT}+1)$ is essentially a
step function at $E_F$ (here $E_F$ is put at zero).
The simple nearly free-electron (NFE) model, with the periodic
potential, leads to a gap $E_g = 2 V_0$ in the free electron band at a
new "zone-boundary" $k=G/2$ \cite{zim,tj1}, and 
the general band dispersion as function of $k$
is 
\begin{equation}
\varepsilon = \frac{1}{2}(k^2+(k-G)^2 \pm \sqrt{(k^2-(k-G)^2)^2+4V_0^2})
\end{equation}
If the gap 
appears at $E_F$ for this particular value of $k$ there is a gain in kinetic energy, $E$.

 For band energies near the gap it is convenient to express the k-dispersion
 in terms of a linear $\epsilon$ measured from $E_F=0$, so that 
 $\epsilon = const. \cdot \kappa$, where $\kappa$ is measured from
 the zone boundary at $G/2$. The approximation of a linear $\epsilon$ as function of $\kappa$
 is valid for $\epsilon << W$, the band width from the bottom
 of the band to $E_F$. Thus,
 
 \begin{equation}
\varepsilon = \pm \sqrt{(\epsilon^2+V_0^2}
\end{equation}

The normal free electron dispersion, $\varepsilon = \epsilon$,
is recovered for $V_0 = 0$, and $N(\varepsilon) = N/|d\varepsilon/d\epsilon|$,
becomes constant and equal to $N$.
The gapped $\Tilde{N}$ is zero for energies within $\pm V_0$ around $E_F$.

The electron-phonon coupling $\lambda$ will enhance the electronic DOS for energies $\pm \hbar \omega$
around $E_F$, where it can be written $N M^2/K$ \cite{zim}.
The matrix
element  $M$ is zero outside the interval $\pm \hbar \omega$.
Inside the interval it can be evaluated as $\langle \Psi^*(E_F,r) dV(r)/du \Psi(E_F,r) \rangle$,
which is the first order change in energy caused by the perturbation $dV(r)$ for $du \rightarrow 0$.
For a finite value of $u$ the change in energy will be finite and equal to the gap $V_0$, 
since $V_0/u$ is constant for harmonic vibrations. 
Thus, instead of calculating $M$ as a matrix element it is possible to take the value directly
from the band gap, and $M$ can be written $V_0/u$ for energies close to $E_F$. The separate values of $V_0$
and $u$ are important variables, and
later it can be verified that the band gap $V_0$ is linked
to the SC gap  $\Delta$.

The energy of an atomic oscillation consists of elastic and
kinetic contributions, but its time dependence,
$2U(t) = K u^2 cos^2(\omega t) + M_A \omega^2 sin^2(\omega t)$,
is a constant in the harmonic approximation. For $t=0$ all the energy is of elastic origin
so the cost in energy to generate an atomic vibration can be written
$U=\frac{1}{2}Ku^2$, where $u$ refers to the maximal atomic displacement. 
Totally there will be a gain in energy if $|U| \leq |E|$. The system will
spontaneously generate phonons in such a case, and this is the condition for the SC state. 
Both $U$ and $E$ are the energies per unit cell
for which the DOS of the normal state is $N$. The condition $|U|=|E|$ is written

\begin{equation}
\frac{1}{2}Ku^2 = \int_{-\hbar\omega}^0 \epsilon (N(\epsilon)-\Tilde{N}(\epsilon)) d\epsilon 
\label{eqn4}
\end{equation}

for $T=0$, where $\Tilde{N}(\epsilon)$ is the DOS with the gap and ${N}(\epsilon)$, the DOS of the
normal state, is assumed
constant within $\hbar \omega$ around $E_F$. The integration is only to $\hbar \omega$ since 
$\lambda$, which will appear later in the equation, is zero for energies larger than $\pm \hbar \omega$.

 With a
substitution $e^2=\epsilon^2+\Delta^2$ we obtain $\Tilde{N}=N|e|/\sqrt{e^2-V_0^2}$ and,

\begin{equation}
\frac{1}{2}Ku^2 = \int_{-\hbar\omega}^0 N \epsilon d\epsilon - \int_{-\hbar\omega}^{-\Delta} Ne^2/\sqrt{(e^2-V_0^2)} de  
\end{equation}

With $\Delta$ replacing $V_0$ this gives
\begin{equation}
Ku^2 = N(\hbar \omega)^2 + N \Delta^2 ln (2\hbar \omega/ \Delta) - N (\hbar \omega)^2 
\end{equation}

and

\begin{equation}
\Delta = 2 \hbar \omega e^{-1/\lambda} 
\end{equation}

since $\lambda = N \Delta^2 / Ku^2$ \cite{tjfe} when the
gap at $T=0$ is $\Delta$.

Therefore, on one hand it can be argued that $V_0$ has to be equal to $\Delta$, the SC gap, 
since this derivation then reproduces the BCS result. But the equivalence between $\Delta$ and
$V_0$ can also be understood from the fact that $V_0$
is a measurable band gap in the superconductor.
Further, a constant
$\lambda$ implies that the phonon amplitude $u$ is proportional to $\Delta$, so that
$u$ is largest at $T$=0, and 
$u \rightarrow 0$ when $\Delta \rightarrow 0$ at $T \rightarrow T_C$.

The integral and the interpretations look a bit different from BCS \cite{bcs,fet},
but the final result for $\Delta$ is the same. 
Any system will generate
phonons as soon as the gain in electronic energy generated by the phonon is larger than the
energy needed for the phonon.  In reality other more subtle effects, electron-electron
correlation energy and potential terms will be added to the energy costs 
and those terms will prevent
a SC gap in many systems. 

A numerical solution needs some care for the diverging part of $\Tilde{N}$, but 
models with non-constant $N$ and with larger ratios of  $\frac{\Delta}{\hbar \omega}$
can be studied.

\section{The limit $\Delta \rightarrow$0.}

The model for finding $T_C$ is obtained from eq. \ref{eqn4}, but with
the Fermi-Dirac function
as the $T$-dependent weight factor for $\Tilde{N}(\varepsilon)$
and $N(\varepsilon)$, and with the integration in the interval [$-\hbar\omega,\hbar\omega$].
Thus, at $T_C$ it is required that the phonon energy is equal to the difference between
the kinetic energy of the gapped and the normal electronic DOS, but with the constraint that
$\Delta \rightarrow$ 0. This is solved numerically from

\begin{equation}
Ku^2 \approx N  I(\hbar\omega,\Delta,T)
\label{eqn8} 
\end{equation}

or

\begin{equation}
1/\lambda \approx  I(\hbar\omega,\Delta,T) /\Delta^2
\label{eqn9}
\end{equation}

where

\begin{equation}
I= \int_{-\hbar\omega}^{\hbar\omega}  \epsilon f d\epsilon -
\int_{-\hbar\omega'}^{\hbar\omega'} e |e|/\sqrt{(e^2-\Delta^2)} f de 
\label{eqn10}
\end{equation}

for $\Delta \rightarrow$ 0. The ' in the first integral means that the energies where $|e| < \Delta$ are excluded.
A factor of $\Delta^2$ is extracted from the right hand side of eq. \ref{eqn8} and combined with 
$N$ and $Ku^2$ on the left hand side to give $\lambda$ as before.

\begin{figure}
\includegraphics[height=6.0cm,width=8.0cm]{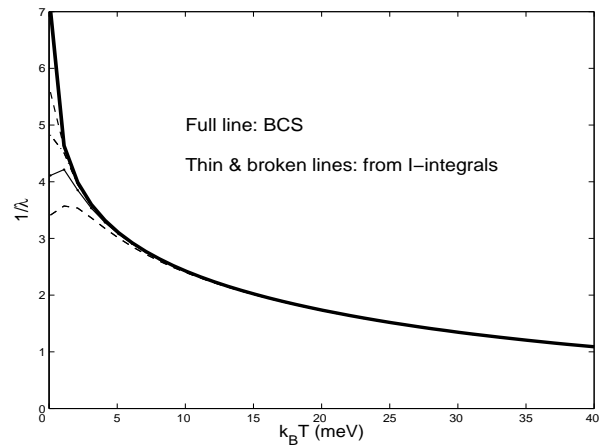}
\caption{The full line shows $1/\lambda$ as function of $k_BT$
as obtained for eq. \ref{eqn11}, i.e. the numerical BCS result for $\hbar\omega$=0.1 eV. The thin and broken
lines show the corresponding result from eqs. \ref{eqn9}-\ref{eqn10}, for 5 different $\Delta$'s
(0.5, 1, 2, 4 and 8 meV). The result for the smallest $\Delta$ is indistinguishable
from the BCS result over this temperature range.
}
\label{fig1}
\end{figure}

The original BCS expression for $T_C$, which is derived from
\begin{equation}
1/gN = \int_0^{\hbar\omega} \frac{de}{e} tanh(e/2k_BT)
\label{eqn11}
\end{equation}
(where $gN$ is the coupling constant \cite{fet}) is independent of $\Delta$. Here, we solve 
this equation numerically with the same
precision as the one-particle expression for very small $\Delta$. The results, shown
in Fig. 1, tend towards the analytic solution of eq. \ref{eqn11}, the well-known BCS formula, which can be written

\begin{equation}
k_BT_C = 1.13 \hbar\omega e^{-1/\lambda}
\end{equation}

when $gN=\lambda$. For example, it can be verified from this formula and Fig \ref{fig1} that a $\lambda$ of 0.5 makes
$k_BT_C \approx$ 15 meV when $\hbar\omega$ is 100 meV.

A non-constant DOS is considered by solving eq. \ref{eqn9}-\ref{eqn10} with the $N(\varepsilon)$ directly from eq. 2
with $\hbar\omega$ being a large fraction of $W$. The requirement of constant number of electrons
as function of $T$ in the normal and SC state makes the numerical stability more difficult. 
But the results indicate
that $T_C$ is lower if the non-constant (free electron like) DOS is included in the integrations
over $\pm \hbar\omega$ compared to the result with a constant DOS. 

For AFM spin waves there is a cost in magnetic energy, which in the harmonic approximation
can be written $U_m = \frac{1}{2}K_m m^2$. The change in potential on some site, $V_m$, is
positive for one spin and negative for the other spin direction. This defines a
$\lambda_{sf} = N V_m^2/K m^2$ as a coupling constant for spin-fluctuations \cite{tjfe}.
The rest of the equations are valid with $\lambda_{sf}$ replacing $\lambda$ and
with $\hbar\omega_{sf}$ being the energy of the spin wave.

\section{Spin-phonon coupling.}

As was mentioned above, typical atomic displacements and magnetic moments from phonons and spin waves in the
normal state can be determined from thermal excitations via the effective force constants
$K$ and $K_m$. 
Magnetic moments, with a tendency for a pseudogap in the DOS, are driven by thermal excitations,
but the left hand side of eq. \ref{eqn8} is larger than the right hand side. However, the situation
might be reversed at lower $T$, when $m$ is supported by the SC gap.
In the SC
state $u$ and $m$ are proportional to the SC gap;
$u=\sqrt(N/K\lambda)\Delta$ and $m=\sqrt(N/K_m\lambda_{sf})\Delta$, respectively.
Increasing amplitudes of $u$ in superconductors at low $T$
should be measurable, but their values are small in conventional superconductors. 
For example, in Nb, with $\lambda \approx 1.2$, $N \approx 0.7 (eV atom spin)^{-1}$,
$K \approx 6 eV/\AA^2$
\cite{papa} and $\Delta \approx 3 meV$, $u/a_0$ ($a_0$ the lattice constant)
will be less than one order
of magnitude smaller than $u$ from zero-point motion. The complex FS of d-band superconductors
implies a multitude of active $q$-vectors. Lithium, which under pressure can have a high $T_C$
and a simple free electron FS, should be more promising for detection of $u$ for a few $q$-vectors. 

Copper oxides with high $T_C$ and relatively simple 2-dimensional FS, should have sizable
amplitudes of the magnetic moments (assuming that spin fluctuations are responsible for superconductivity),
and simple q-dependence.
Both the superconducting gap and the pseudogap for $T>T_C$ are sensitive to
spin waves (or phonons) with the same $q$-vectors. For instance,
fluctuations in form of spin waves and SPC in the cuprates 
are thermally excited at large $T$ and contribute to a pseudogap
for $T \leq T^*$ \cite{tj7}. 
The fact that $m$ (or $u$) is proportional to $\Delta$ shows that these magnetic fluctuations will
reappear in the SC state and become stronger as $T \rightarrow 0$.  
This is in line with the observations of increasing peak intensity 
of spin waves at or below $T_C$ in experiments of inelastic neutron scattering
on underdoped YBCO \cite{hin}. Theoretical estimates of $m$, in the range 0.1-0.2 $\mu_B/Cu$
in the SC state and in the normal state at large $T$, are not very precise because of uncertainties in
density functional calculations.
However, the q-dependence is not expected to change as $T$ goes below $T_C$, so the
results for spin excitations calculated for the normal state in ref. \cite{tj9}
can be carried over to the SC state.

The standard propositions for higher $T_C$ is to increase $\omega$ 
(through isotope shifts) and/or $\lambda$
(through higher absolute value of the DOS \cite{apl}). It is seen that $\lambda$
can remain constant and lead to larger $\Delta$ if $m$ (or $u$) is increased.
This might be achieved through anharmonicity so the maximum $m$ are increased 
without large changes of $K_m$.
Another possibility is to modify the way $k_BT$ is occupying and depleting states
around the SC gap and around $E_F$ in the normal state. Temperature and the Fermi-Dirac
distribution cannot change, but different energy variations of the DOS near $E_F$ are
possible. The effect of a gap originating from a potential perturbation on the
real band structure has to be determined and inserted in eqs. \ref{eqn9}-\ref{eqn10} for sorting out
the effects of non-constant DOS, $N(\varepsilon)$.

The case with strong SPC in the cuprates leads to large enhancements of $\lambda_{sf}$ for some
phonons with particular $q$-vectors. Anharmonicity is also expected in SPC when the magnetic
moments can be enhanced at large $u$, which then lead to mutual softening of phonons and
spin waves. This can be taken into account in quasi-harmonic vibrations as amplitude dependent
$K_m$ and $K$. When
the gap at $E_F$ is caused by a few
periodicities of phonons or spin waves it might be important to consider
decoupled $q$-dependencies in the numerical search of $\Delta$ and $T_C$.
For instance, large $\lambda_{SPC}$ for selective $\vec{q}$ on the 2-dimensional
FS of the high-$T_C$ cuprates is favorable to d-wave pairing, because the strength 
of the gap is different at different parts of the FS 
 \cite{tj6,sjpw}. However, the present derivation does not yet consider the phase of the gap.  
In direct SPC, the excitations of a phonon and a spin wave are made together,
which can make $\omega_{sf} \approx \omega$ and a
large $\lambda_{SPC}$. Indirect, weaker $\lambda_{sf}$ are possible for independent spin fluctuations
which are enhanced by the presence of normal thermally excited phonons at large T \cite{acmp,tj9}.
This relative strength of SPC is consistent with
effective couplings derived from optical spectra showing that the strongest couplings appear
near phonon energies, while weaker couplings exist at higher energies \cite{heum}.

\section{Field dependence.}

From the discussion about selective q-dependence there is a possible reason
to why a weak magnetic field is expelled in a superconductor. As mentioned, one
particular phonon, $\vec{q}$, is responsible for the gap on
the paramagnetic FS at $E_F$. A magnetic field, $H$, will split the FS into two, one for each
spin ("up" or "down"), $E_F^{u,d} = E_F \pm \mu_BH$. If SC gaps
should appear optimally on these two FS, it requires two independent phonons.
One produces a potential perturbation at $e^{-i(\vec{q}-\vec{\delta})\cdot \vec{x}}$,
and the other at $e^{-i(\vec{q}+\vec{\delta})\cdot \vec{x}}$, where $\delta$ is
determined by the band dispersion and $\mu_BH$. The sum of these two potentials
is $2cos(\vec{\delta}\cdot \vec{x})e^{-i\vec{q}\cdot \vec{x}}$, and therefore,
even if there is a modulation given by the cosine function,
the effective $\vec{q}$ remains the same and will not fit to the optimal values
for the two FS.
The resulting SC gaps do not appear at $E_F$ on the two spin-split bands.
This will reduce the gain in energy of the SC state, as can be seen
from a two-level model:   
Suppose that the DOS of the SC state consists of two peaks at $\pm \Delta$ (each containing one
electron of each spin) and that $E_F$=0 in zero field.
The kinetic energy in this case 
\begin{equation}
E_0 = 2(f(-\Delta,T)(-\Delta )+ f(\Delta,T)(\Delta))
\end{equation}
A field puts the states asymmetrically around $E_F$ and the kinetic energy
for the two spin states ("up" or "down") will be
\begin{equation}
E_H^{\pm}= f(-\Delta \pm h,T)(-\Delta \pm h) +  f(\Delta \pm h,T)(\Delta \pm h)
\end{equation}
where $h=\mu_BH$.
The result is that $E_0 < E_H^-+E_H^+$ for most $T>0$ (but for $k_BT$ and $h$ being small
in comparison to $\Delta$), i.e. the symmetric state with no field
has the lowest energy. 

\begin{figure}
\includegraphics[height=6.0cm,width=8.0cm]{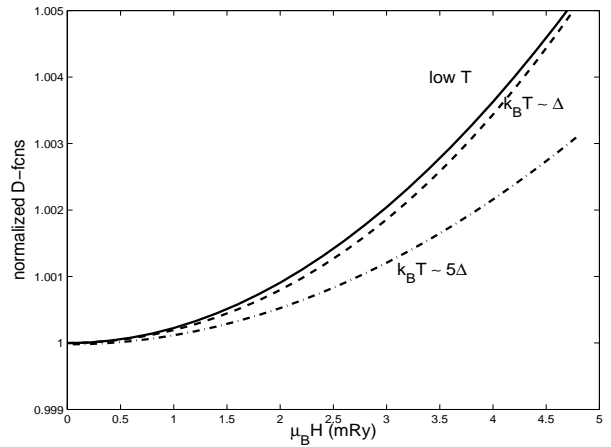}
\caption{The full, broken and semi-broken lines show
the values of the energy integral $D(h,T)$ (see eq. \ref{eqnD}) as function of $h$
for low, intermediate, and high temperature,
respectively. The values are normalized to 1 for $h=0$.  These calculations are made with
$\Delta=5 mRy$ and $\hbar \omega = 50 mRy$.
}
\label{fig2}
\end{figure}

The increasing kinetic energy can also be demonstrated for the approximation of a constant DOS by adding
and subtracting the field $h$ in the Fermi-Dirac function. 
Figure \ref{fig2} shows the energy difference, $D(h,T)$, of kinetic energy for the gapped superconducting DOS
with and without field, which is calculated as:

\begin{equation}
D(h,T)=\int_{-\hbar\omega'}^{\hbar\omega'}e\Tilde{N}(e)(f(e+h,T)+f(e-h,T)-2f(e,T))de 
\label{eqnD}
\end{equation}

when $h < \Delta$ for low and high $T$ ($\approx \Delta$). The increase of $D$ as function of the field
$h$ is because the thermal occupation can be made more efficiently if $E_F$ is closer to the DOS peak (on $\Tilde{N}$) 
above the gap for "majority" and closer to the DOS peak below the gap in the "minority" states,
than if $E_F$ is in the middle of the gap. 
Thus, the gapped state with $\mu_BH$=0 has the lowest kinetic energy.
This state will be preferred by the system
as long as screening of an external field can be made through superconducting currents.
The model shows that the minimum at $h=0$ is less profound for large $T$, or when $k_BT$ exceeds
about $5\Delta$. 
The feedback from the transfer of minority to majority spin
states and effects of a non-constant DOS are not included in the model.

\section{Conclusion.}

The BCS formulas for $\Delta$ at $T=0$ and $T_C$ at $\Delta=0$ are derived
directly from the one-particle DOS functions of the gapped and normal state band structures.
This allows for an easy comprehension
and further interpretations of the SC mechanism. While 
phonons and/or spin waves are excited thermally in the normal state, they are generated 
via the electronic band gap in
the superconducting state. Atomic displacements of harmonic vibrations and magnetic moments
of harmonic spin fluctuations are proportional to the SC gap. 
Since the SC gap is closely related to the gap of the perturbed band structure, it will be interesting to consider
DOS functions in materials with impurities and other defects via supercell calculations.

\end{document}